\documentstyle[aps,prd,preprint]{revtex}
\begin{document}
\draft
\title{Acoustic propagation in fluids:\\
       an unexpected example of Lorentzian geometry}
\author{Matt Visser\cite{e-mail}}
\address{Physics Department, Washington University, St. Louis,
         Missouri 63130-4899}
\date{11 October 1993}
\maketitle
\begin{abstract}
It is a deceptively simple question to ask how acoustic disturbances
propagate in a non--homogeneous flowing fluid.  If the fluid is
barotropic and inviscid, and the flow is irrotational (though it
may have an arbitrary time dependence), then the equation of motion
for the velocity potential describing a sound wave can be put in
the $(3+1)$--dimensional form $\Delta \psi \equiv \partial_\mu
\left( \sqrt{-g}\; g^{\mu\nu} \; \partial_\nu \psi \right)/\sqrt{-g}
= 0$.  The {\sl acoustic metric} $g_{\mu\nu}(t,\vec x)$ governing
the propagation of sound  depends on the density, flow velocity,
and local speed of sound. Even though the underlying fluid dynamics
is Newtonian, non--relativistic, and takes place in flat space +
time, the fluctuations (sound waves) are governed by a Lorentzian
spacetime geometry.
\end{abstract}

\narrowtext
\section{INTRODUCTION}

It is well known that for a static homogeneous inviscid fluid the
propagation of sound waves is governed by the simple equation
\begin{equation}
\partial_t^2 \psi = c^2 \nabla^2 \psi.
\end{equation}
($c \equiv \hbox{speed of sound}$.) It is a deceptively simple
question to ask what happens if the fluid is non--homogeneous, in
motion, possibly even in non-steady motion.  If the fluid is
barotropic and inviscid, and the flow is irrotational (though it
may have an arbitrary time dependence) then I shall show that the
equation of motion for the velocity potential describing an acoustic
disturbance  can be put in the $(3+1)$--dimensional form
\begin{equation}
\Delta \psi \equiv
{1\over\sqrt{-g}}
\partial_\mu \left( \sqrt{-g} \; g^{\mu\nu} \; \partial_\nu \psi \right) = 0.
\end{equation}
The propagation of sound is governed by the {\sl acoustic metric}
$g_{\mu\nu}(t,\vec x)$. This acoustic metric describes a Lorentzian
(pseudo--Riemannian) geometry and depends on the density, velocity
of flow, and local speed of sound in the fluid. Specifically
\begin{equation}
g_{\mu\nu}(t,\vec x)
\equiv {\rho\over c}
\left[ \matrix{-(c^2-v^2)&\vdots&-{\vec v}\cr
               \cdots\cdots\cdots\cdots&\cdot&\cdots\cdots\cr
	       -{\vec v}&\vdots&\tensor I\cr } \right]_.
\end{equation}
In general, when the fluid is non--homogeneous and flowing, the
{\em acoustic Riemann tensor} associated with this Lorentzian metric
will be nonzero.  It is quite remarkable that even though the
underlying fluid dynamics is Newtonian, nonrelativistic, and takes
place in flat space + time, the fluctuations (sound waves) are
governed by a curved Lorentzian (pseudo--Riemannian) geometry.
This connection between fluid dynamics and techniques more commonly
encountered in the context of general relativity opens up many
opportunities for cross--pollination between the two fields.

\section{FLUID DYNAMICS}

\subsection{Fundamental equations}

The fundamental equations of fluid
dynamics~\cite{Lamb,Landau-Lifshitz,Milne-Thomson} are
the equation of continuity
\begin{equation}
\partial_t \rho + \nabla\cdot(\rho \vec v) = 0,
\end{equation}
and Euler's equation
\begin{equation}
\rho {d \vec v\over dt} \equiv
\rho \left( \partial_t \vec v + (\vec v \cdot \nabla) \vec v \right) =
\vec F.
\end{equation}
Start by assuming the fluid to be inviscid, with the only forces
present being those due to pressure and Newtonian gravity
\begin{equation}
\vec F = - \nabla p - \rho \nabla \phi.
\end{equation}
Standard manipulations yield
\begin{equation}
\partial_t {\vec v} =
\vec v \times ( \nabla \times \vec v)  - {1\over\rho} \nabla p
- \nabla( {\scriptstyle{1\over2}} v^2 + \phi).
\end{equation}
Take the flow to be irrotational, introducing the velocity potential
$\psi$ such that $\vec v = -\nabla \psi$. Take the fluid to be
barotropic ($\rho$ is a function of $p$ only). Then define
\begin{equation}
\zeta(p) = \int_0^p {dp'\over\rho(p')};
\qquad \hbox{so that} \qquad
\nabla \zeta = {1\over\rho} \nabla p.
\end{equation}
Euler's equation reduces to
\begin{equation}
-\partial_t \psi + \zeta + {\scriptstyle{1\over2}} (\nabla\psi)^2
+ \phi = 0.
\end{equation}

\subsection{Fluctuations}

Linearize the equations of motion around some assumed background
$(\rho_0,p_0,\psi_0)$ by setting $\rho = \rho_0 + \epsilon \rho_1$,
$p = p_0 + \epsilon p_1$, and $\psi = \psi_0 + \epsilon \psi_1$.
The gravitational potential $\phi$ is taken to be fixed and external.
Sound is {\em defined} to be these linearized fluctuations in the
dynamical quantities.  The linearized continuity equation reads
\begin{equation}
\partial_t \rho_1 + \nabla\cdot(\rho_1 \vec v_0 + \rho_0 \vec v_1) = 0,
\end{equation}
while from the Euler equation,
using $\zeta = \zeta_0 + \epsilon (p_1/\rho_0)$,
\begin{equation}
-\partial_t \psi_1 + {p_1\over\rho_0} - \vec v_0 \cdot \nabla\psi_1 = 0.
\end{equation}
Rearranging
\begin{equation}
p_1 =  \rho_0 ( \partial_t \psi_1 + \vec v_0 \cdot \nabla\psi_1 ).
\end{equation}
Substitute this linearized Euler equation into the linearized
equation of continuity. Use $\rho_1 = (\partial \rho/\partial p)
p_1$. One obtains, up to an overall sign,
\begin{equation}
- \partial_t
     \left( {\partial\rho\over\partial p} \rho_0 \;
            ( \partial_t \psi_1 + \vec v_0 \cdot \nabla\psi_1 )
     \right)
+ \nabla \cdot
     \left( \rho_0 \nabla\psi_1
            - {\partial\rho\over\partial p} \rho_0 \vec v_0 \;
	      ( \partial_t \psi_1 + \vec v_0 \cdot \nabla\psi_1 )
     \right)
=0.
\end{equation}
This wave equation describes the propagation of the scalar potential
$\psi_1$, and thereby completely determines the quantities $p_1$
and $\rho_1$. The background fields $p_0$, $\rho_0$ and $\vec v_0$
are permitted to have {\em arbitrary} temporal and spatial dependencies.
Now, written in this form, the physical import of the wave equation
is somewhat less than pellucid. Define $1/c^2 = \partial\rho/\partial
p$, and construct the symmetric $4\times4$ matrix
\begin{equation}
f^{\mu\nu} \equiv
{\rho_0\over c^2}
\left[ \matrix{-1&\vdots&-v_0^j\cr
               \cdots\cdots&\cdot&\cdots\cdots\cdots\cdots\cr
	       -v_0^i&\vdots&( c^2\delta^{ij} - v_0^i v_0^j )\cr }
\right]_.
\end{equation}
Then, using four dimensional coordinates $x^\mu = (t, x^i)$ the
wave equation is easily rewritten as
\begin{equation}
\partial_\mu ( f^{\mu\nu} \partial_\nu \psi_1) = 0.
\end{equation}
This remarkably compact formulation is much more promising. The
remaining steps are a straightforward application of the techniques
of curved space Lorentzian geometry.

\section{LORENTZIAN GEOMETRY}

In any Lorentzian (pseudo--Riemannian) manifold the curved space
scalar d'Alembertian is given in terms of the metric $g_{\mu\nu}(t,\vec
x)$ by~\cite{Fock,Moller,MTW,Hawking-Ellis}
\begin{equation}
\Delta \psi \equiv
{1\over\sqrt{-g}}
\partial_\mu \left( \sqrt{-g} \; g^{\mu\nu} \; \partial_\nu \psi \right).
\end{equation}
The inverse metric, $g^{\mu\nu}(t,\vec x)$ is pointwise the matrix
inverse of $g_{\mu\nu}(t,\vec x)$, while $g \equiv \det(g_{\mu\nu})$.
Thus we can rewrite our physically derived wave equation in terms
of the d'Alembertian provided we identify
\begin{equation}
\sqrt{-g} \; g^{\mu\nu} = f^{\mu\nu}.
\end{equation}
This implies
\begin{eqnarray}
g  &=& \det(f^{\mu\nu})
\nonumber\\
&=& \left({\rho_0/c^2}\right)^4
   [-1 \cdot (c^2 - v_0^2) - v_0^2] (c^2)^2 = - \rho_0^4/c^2.
\end{eqnarray}
We can thus pick off the coefficients of the inverse metric
\begin{equation}
g^{\mu\nu} \equiv
{1\over \rho_0 c}
\left[ \matrix{-1&\vdots&-v_0^j\cr
               \cdots\cdots&\cdot&\cdots\cdots\cdots\cdots\cr
	       -v_0^i&\vdots&(c^2 \delta^{ij} - v_0^i v_0^j )\cr }
\right]_.
\end{equation}
One could now determine the metric itself by inverting this $4\times4$
matrix. On the other hand it is even easier to recognize that we
have in front of us an example of the Arnowitt--Deser--Misner split
of a $(3+1)$--dimensional  Lorentzian spacetime metric into space
+ time, more commonly used in discussing the initial value data in
general relativity (see, for example, \cite{MTW} pp 505--508). The
metric is
\begin{equation}
g_{\mu\nu} \equiv
{\rho_0 \over  c}
\left[ \matrix{-(c^2-v_0^2)&\vdots&-v_0^j\cr
               \cdots\cdots\cdots\cdots&\cdot&\cdots\cdots\cr
	       -v_0^i&\vdots&\delta^{ij}\cr }
\right]_.
\end{equation}
Observe that the signature of this metric is in fact $(-,+,+,+)$, as
it should be.

It should be emphasised that there are two distinct metrics relevant
to the current discussion. The {\em physical spacetime metric} is
just the usual flat metric of Minkowski space $\eta_{\mu\nu} \equiv ({\rm
diag} [-c_\infty^2,1,1,1])_{\mu\nu}$.  ($c_\infty = \hbox{speed of
light}$.) The fluid particles couple only to the physical metric
$\eta_{\mu\nu}$. In fact the fluid motion is completely non--relativistic
--- $||v_0|| \ll c_\infty$. Sound waves on the other hand, do not
``see'' the physical metric at all. Acoustic perturbations couple
only to the {\em acoustic metric} $g_{\mu\nu}$.  The geometry
determined by the acoustic metric does however inherit some key
properties from the existence of the underlying flat physical
metric.

For instance, the topology of the manifold does not depend on the
particular metric considered.  The acoustic geometry inherits
the underlying topology of the physical metric --- $\Re^4$ with
possibly a few regions excised (due to imposed boundary conditions).

Furthermore the acoustic geometry  automatically inherits the
property of ``stable causality''~\cite{Hawking-Ellis}. Note that
$g^{\mu\nu} (\nabla_\mu t) (\nabla_\nu t) = -1/(\rho_0 c) < 0$.
This precludes some of the more entertaining pathologies that
sometimes arise in general relativity.

Another concept that translates immediately is that of an
``ergo--region''. Consider integral curves of the vector
$(\partial/\partial t)^\mu \equiv (1,0,0,0)^\mu$. Then $g_{\mu\nu}
(\partial/\partial t)^\mu (\partial/\partial t)^\nu = g_{tt} =
-[c^2 - v_0^2]$.  This changes sign when $||\vec v_0|| > c$. Thus
any region of supersonic flow is an ergo--region.  The analogue of
this behaviour in general relativity is the ergosphere surrounding
any spinning black hole --- it is a region where space ``moves''
with superluminal velocity relative to the fixed stars.

Observe that in a completely general Lorentzian geometry the metric
has 6 degrees of freedom per point in spacetime. ($4\times4$
symmetric matrix $\Rightarrow$ $10$ independent components; then subtract
$4$ coordinate conditions).  In contrast, the acoustic metric is
more constrained. Being specified completely by the three scalars
$\psi_0(t, \vec x)$, $\rho_0(t, \vec x)$, and $c(t, \vec x)$, the
acoustic metric has only $3$ degrees of freedom per point in
spacetime.

A point of notation: Where the general relativist uses the word
``stationary'' the fluid dynamicist uses the phrase ``steady flow''.
Where the general relativist uses the word ``static'' the fluid
mechanic would translate this as ``fluid at rest''.

The analogies I am invoking between acoustics in fluids and general
relativity are very deep and very powerful --- there is a lot of
mathematical machinery  available for use.

\section{GEOMETRIC ACOUSTICS}

Taking the short wavelength/high frequency limit to obtain geometrical
acoustics is now easy. Sound rays (phonons) follow the {\em null
geodesics} of the acoustic metric. Compare this to general relativity
where in the geometrical optics approximation light rays (photons)
follow {\em null geodesics} of the physical spacetime metric. Since
null geodesics are insensitive to any overall conformal factor in
the metric~\cite{MTW,Hawking-Ellis} one might as well simplify life
by considering the metric
\begin{equation}
h_{\mu\nu} \equiv
\left[ \matrix{-(c^2-v_0^2)&\vdots&-v_0^j\cr
               \cdots\cdots\cdots\cdots&\cdot&\cdots\cdots\cr
	       -v_0^i&\vdots&\delta^{ij}\cr }
\right]_.
\end{equation}
Thus, in the geometric acoustics limit, sound propagation is
insensitive to the density of the fluid, and depends only on the
local speed of sound and the velocity of the fluid. It is only for
specifically wave related properties that the density of the medium
becomes important.

One can rephrase this in a language more familiar to the acoustics
community. Take $\psi_1 \sim a e^{i\varphi}$. Then, neglecting
variations in the amplitude $a$, the wave equation reduces to the
{\em Eikonal equation}
\begin{equation}
h^{\mu\nu} \; \partial_\mu \varphi \; \partial_\nu \varphi = 0.
\end{equation}
This Eikonal equation is blatantly insensitive to any overall
multiplicative prefactor.

As a sanity check on the formalism, let the null geodesic be
parameterized by $X^\mu(t) \equiv (t, \vec x(t))$. Then the null
condition implies
\begin{eqnarray}
&& h_{\mu\nu} {dX^\mu \over dt} {dX^\nu \over dt} = 0
\nonumber\\
&&\iff
-(c^2 - v_0^2) - 2 v_0^i {dx^i \over dt}
+ {dx^i \over dt} {dx^i \over dt} = 0
\nonumber\\
&&\iff
\left\Vert {d{\vec x} \over dt} - \vec v_0 \right\Vert = c.
\end{eqnarray}
Here the norm is taken in the flat physical metric. This has the
obvious interpretation that the ray travels at the speed of sound
relative to the moving medium.

If the geometry is stationary, one can do slightly better. Let
$X^\mu(s) \equiv (t(s); \vec x(s))$ be some null path from $\vec x_1$
to $\vec x_2$ parameterized in terms of physical arc length ({\em
i.e.} $|| d{\vec x}/ds || \equiv 1$). Then the condition for the
path to be null (though not yet necessarily a null geodesic) is
\begin{equation}
-(c^2 - v_0^2) \left({dt\over ds}\right)^2
- 2 v_0^i \left({dx^i\over ds}\right) \left({dt\over ds}\right)
+1 = 0.
\end{equation}
Solving the quadratic
\begin{equation}
\left({dt\over ds}\right)
= { - v_0^i \left({dx^i\over ds}\right)
    + \sqrt{ c^2 - v_0^2 + \left(v_0^i {dx^i\over ds}\right)^2 }
   \over
   c^2 - v_0^2}.
\end{equation}
The total time taken to traverse the path is thus
\begin{eqnarray}
T[\gamma]
&=& \int_{\vec x_1}^{\vec x_2} (dt/ds) ds
\nonumber\\
&=& \int_\gamma \{
                 \sqrt{ (c^2 - v_0^2)ds^2 + (v_0^i dx^i)^2 }
		 - v_0^i dx^i
                \} / (c^2-v_0^2).
\end{eqnarray}
Extremizing the total time taken {\em is} Fermat's principle for
sound rays. One has thus checked the formalism for stationary
geometries (steady flow) by reproducing the discussion on p 262 of
Landau and Lifshitz~\cite{Landau-Lifshitz}.

\section{DISCUSSION}
\subsection{Limitations}

The derivation of the wave equation made two key assumptions ---
the flow is irrotational flow and the fluid is barotropic.

The d'Alembertian equation of motion for acoustic disturbances,
though derived only under the assumption of irrotational flow,
continues to make perfectly good sense in its own right if the
background velocity field $\vec v_0$ is given some vorticity.
This leads one to hope that it {\em might} be possible to find a
suitable generalization of the present derivation that might work
for flows with nonzero vorticity. In this regard, note that if the
vorticity is everywhere confined to thin vortex filaments, the
present derivation already works everywhere outside the vortex
filaments themselves.

The restriction to a barotropic fluid ($\rho$ a function of $p$
only) is in fact related to issues of vorticity. Examples of
barotropic fluids are:\hfil\break
$\bullet$ Isothermal fluids subject to isothermal perturbations.\hfil\break
$\bullet$ Fluids in convective equilibrium subject to adiabatic
perturbations.\hfil\break
See for example \cite{Lamb}, \S 311, pp 547--548, and \S 313 pp 554--556.
Failure of the barotropic condition implies that the perturbations
cannot be vorticity free and requires more sophisticated analysis.

\subsection{Precursors}

It is perhaps surprising that anything new can be said about so
venerable a subject  as fluid dynamics. Certainly there are precursors
to the discussion of this letter in the fluid dynamics literature.
For instance, take the background to be static, so that $\vec
v_0=0$, while $\partial_t \rho_0 = 0 = \partial_t p_0$, though
$p_0$ and hence $c$ are permitted to retain arbitrary spatial
dependencies. Then the wave equation derived in this letter reduces to
\begin{equation}
\partial_t^2 \psi = c^2 {1\over \rho_0} \nabla \cdot (\rho_0 \nabla \psi).
\end{equation}
This equation is in fact well known. It is equivalent, for instance
to eq.~(13) of \S 313 of Lamb's classic {\sl Hydrodynamics}~\cite{Lamb}.
See also eq.~(1.4.5) of the recent book by DeSanto~\cite{deSanto}.
The superficially similar wave equations discussed by Landau and
Lifshitz~\cite{Landau-Lifshitz} (see \S 74, eq.~(74.1)), and by
Skudrzyk~\cite{Skudrzyk} (see p 282), utilize different physical
assumptions concerning the behaviour of the fluid.  The novelty I
am claiming for the current letter is firstly, inclusion of nonzero
background velocities and arbitrary time dependencies, and secondly
and more importantly, the interpretation of these results in terms
of Lorentzian geometry.

\section{CONCLUSIONS}

I have shown that acoustic waves in an inviscid fluid can (under
the assumptions of irrotational barotropic flow) be described by
the scalar d'Alembertian of a suitable Lorentzian geometry.  For
inhomogeneous flows this Lorentzian geometry will exhibit nonzero
curvature.

Prior to this observation, Lorentzian geometries have been of interest
to physics only within the confines of Einstein's theory of
gravitation (general relativity). A large quantity of technical
mathematical machinery currently utilized only within the context of
general relativity may thus become of interest to the fluid dynamics
community.

On the other hand, the results of this letter give the general
relativists a very down to earth physical model for certain classes
of Lorentzian geometry.

Particularly intriguing is the fact that while the underlying
physics of fluid dynamics is completely nonrelativistic, Newtonian,
and sharply separates the notions of space and time, one nevertheless
sees that the fluctuations couple to a full--fledged Lorentzian
{\em spacetime}.

{\em Note Added:} After this paper was submitted for publication
I was informed that similar results can be found in the interesting
but little--known work of Unruh~\cite{Unruh}. In that Letter, Unruh
investigated the acoustic equivalent of Hawking radiation arising
from the fluid dynamical analogue of a black hole. I wish to thank
Ted Jacobson and John Friedman for bringing this reference to my
attention.  Further work on acoustic black holes may be found
in~\cite{Jacobson}.

\acknowledgements

This research was supported by the U.S. Department of Energy.


\end{document}